%% file: main.tex
\documentclass[journal]{IEEEtran}
\usepackage[T1]{fontenc}

\usepackage{array}

\usepackage{graphicx} 
\graphicspath{{./figs/}}

\usepackage{subcaption}
\usepackage{amsmath, amssymb, amsfonts}
\usepackage{color}
\usepackage{mathtools}
\usepackage{algorithm}
\usepackage[noend]{algorithmic}

\usepackage{multirow}
\usepackage{float}
\usepackage{url} 
\usepackage{comment}
\usepackage{cite}

\usepackage[acronym]{glossaries}
\loadglsentries[acronym]{./utils/acronyms}

\newlength\myindent
\newlength\myindentt
\usepackage[normalem]{ulem} 
\newcommand{\stkout}[1]{%
  \ifmmode
    \cancel{#1}%
  \else
    \sout{#1}%
  \fi
}

\usepackage{tikz}

\newcommand\copyrighttext{%
  \footnotesize \textcopyright \the\year{} IEEE. Personal use of this material is permitted.  Permission from IEEE must be obtained for all other uses, in any current or future media, including reprinting/republishing this material for advertising or promotional purposes, creating new collective works, for resale or redistribution to servers or lists, or reuse of any copyrighted component of this work in other works.}

\newcommand\copyrightnotice{%
\begin{tikzpicture}[remember picture,overlay]
\node[anchor=south,yshift=5pt] at (current page.south) {\fbox{\parbox{\dimexpr0.90\textwidth-\fboxsep-\fboxrule\relax}{\copyrighttext}}};
\end{tikzpicture}%
}

\begin{document}

\title{ISAC Beamforming Design Based on a Matrix Nearness Formulation with Improved Efficiency}
%
%
%
\author{Berkan Kilic,
        Kenan Turbic,~\IEEEmembership{Member,~IEEE,}
        and
        S\l{}awomir Sta\'nczak,~\IEEEmembership{Senior Member,~IEEE}%
\thanks{%
%
The authors acknowledge the financial support by the Federal Ministry of Research, Technology and Space (BMFTR) in Germany in the programme of ``Souverän. Digital. Vernetzt.'' Joint project 6G-RIC, project identification numbers: 16KISK020K and 16KISK030.
(\textit{Corresponding author: {Berkan Kilic}.})
}%
    \thanks{
    The authors are with the Wireless Communications and Networks Department, Fraunhofer Heinrich Hertz Institute (HHI), 10587, Berlin, Germany (e-mail: \{berkan.kilic, kenan.turbic, slawomir.stanczak\}@hhi.fraunhofer.de).
    }%
    \thanks{S. Sta\'nczak is also with Technische Universit{\"a}t Berlin, 10587, Berlin, Germany.
    }
}

\markboth{}%
{Kilic et al.: ISAC Beamforming Design Based on a Matrix Nearness Formulation with Improved Efficiency}
\maketitle

\IEEEpeerreviewmaketitle

\begin{abstract}
We propose an \gls{isac} beamforming method that performs joint \gls{mimo} radar sensing and \acrlong{mumimo} communication. Our approach builds on a matrix nearness formulation of the \gls{mimo} radar problem and utilizes our recently proposed efficient solver, where the computational complexity is dominated by an \gls{evd} evaluation at each iteration. We extend this formulation to an \gls{isac} scenario by incorporating minimum \acrlong{snr} constraints as communication design criteria, only requiring statistical \acrlong{csi} knowledge at \acrlong{bs}. 
Furthermore, we propose a method to avoid the burdensome \gls{evd} evaluations in certain iterations, reducing the computation time by up to six times in a massive \gls{mimo} setting.   
\end{abstract}


\glsresetall 

\section{Introduction}
\IEEEPARstart{B}{eamforming} plays a critical role in achieving high performance and energy efficiency in wireless multi-antenna communication and radar systems \cite{tse2005fundamentals}.
Its primary objective is to shape the spatial distribution of the transmitted power by concentrating radiation toward the directions of interest, where radar targets or communication users are {expected to be located}, while suppressing radiation elsewhere \cite{gershman2010convex}. This capability is essential for the efficient use of available power and spectral resources \cite{li2008mimo}.

In \gls{isac} systems, designing computationally efficient beamforming techniques is an essential task. A widely adopted approach aims to ensure communication \gls{qos} with minimal degradation in sensing performance \cite{wen2024survey, tang2025dual}. The sensing objective is typically to maximize the effective radar target illumination while minimizing cross-correlations, following the state-of-the-art \gls{mimo} radar framework from \cite{stoica2007probing}. This formulation, augmented with communication \gls{qos} constraints, serves as a foundation in many \gls{isac} beamforming studies \cite{liu2020joint, hua2023optimal}.  

However, the high computational complexity of this approach renders it impractical for massive \gls{mimo} scenarios \cite{globecom_bk, kilic_jsac}.
Another fundamental limitation of many \gls{isac} beamforming methods is their reliance on perfect instantaneous \gls{csi} knowledge at the \gls{tx} \cite{liu2020joint, hua2023optimal}. 
{However, }accurate estimation {of instantaneous \gls{csi}} is particularly difficult in \gls{fdd} systems due to frequency separation between uplink and downlink channels \cite{bengtsson2018optimum}. Moreover, it necessitates frequent feedback from users to \gls{bs}, leading to significantly increased communication overhead. Consequently, more practical and robust {beamforming} approaches assume only statistical \gls{csi} knowledge at the \gls{tx} \cite{huang2025statistical}, leading to the development of long-term beamforming strategies, which need to adapt only to the changes in the large-scale fading parameters \cite{lozano2007long}. 

{In this letter, we present a covariance matrix design approach for long-term \gls{isac} beamforming that extends our previous matrix nearness-based \gls{mimo} radar design method \cite{globecom_bk, kilic_jsac}.}
We propose a novel reformulation of this method, with two major contributions.
First, {we include} minimum required \gls{snr} levels for communication users as additional design constraints, thereby providing a mechanism to incorporate \gls{qos} requirements for reliable \gls{mumimo} communication.
Second, we propose a new solver, leveraging the {Courant-Fischer theorem \cite{marshall1979inequalities}} 
to relax the computational burden by avoiding evaluation of \glspl{evd} in the initially proposed algorithm.
Numerical results demonstrate the ability of the proposed method to design beampatterns that satisfy both communication and sensing requirements, while significantly reducing the computation time required for their synthesis. 

The rest of this letter is structured as follows. Section \ref{Sec:SystemModel} presents the system model, Section \ref{Sec:DesignOptCriteria} defines the adopted communication and sensing design metrics, and Section \ref{Sec:ProposedMethodology} presents the proposed beamforming approach.
Numerical results are presented in Section \ref{Sec:NumResults} and the concluding remarks are given in Section \ref{Sec:Conclusions}.

The following notation is adopted. Matrices and vectors are represented by uppercase and lowercase bold letters, respectively. The \( i \)-th entry of a vector \( \boldsymbol{x} \) is \( x_i \), while \( \boldsymbol{x}_j \) denotes the \( j \)-th column of a matrix \( \boldsymbol{X} \), with \( X_{ij} \) being its \( ij \)-th entry. The \(\ell_2\)-norm of \( \boldsymbol{x} \) is \( \|\boldsymbol{x}\| \), and the Frobenius norm of \( \boldsymbol{X} \) is \( \|\boldsymbol{X}\|_F \). The transpose and Hermitian transpose of \( \boldsymbol{X} \) are \( \boldsymbol{X}^T \) and \( \boldsymbol{X}^H \), respectively. The trace of \( \boldsymbol{X} \) is \( \mathrm{Tr}(\boldsymbol{X}) \), and (\( \boldsymbol{X} \succeq \boldsymbol{0} \)) $\boldsymbol{X} \succ \boldsymbol{0}$ indicates that \( \boldsymbol{X} \) is positive (semi-)definite. The identity matrix of size \( N \) is \( \boldsymbol{I}_N \), and \( \mathbb{E}[\,\cdot\,] \) denotes statistical expectation. 

\vspace{-5mm}
 \copyrightnotice

\section{System Model}
\label{Sec:SystemModel}
\begin{figure}[t]
    \centering
    \includegraphics[width=0.99\columnwidth]{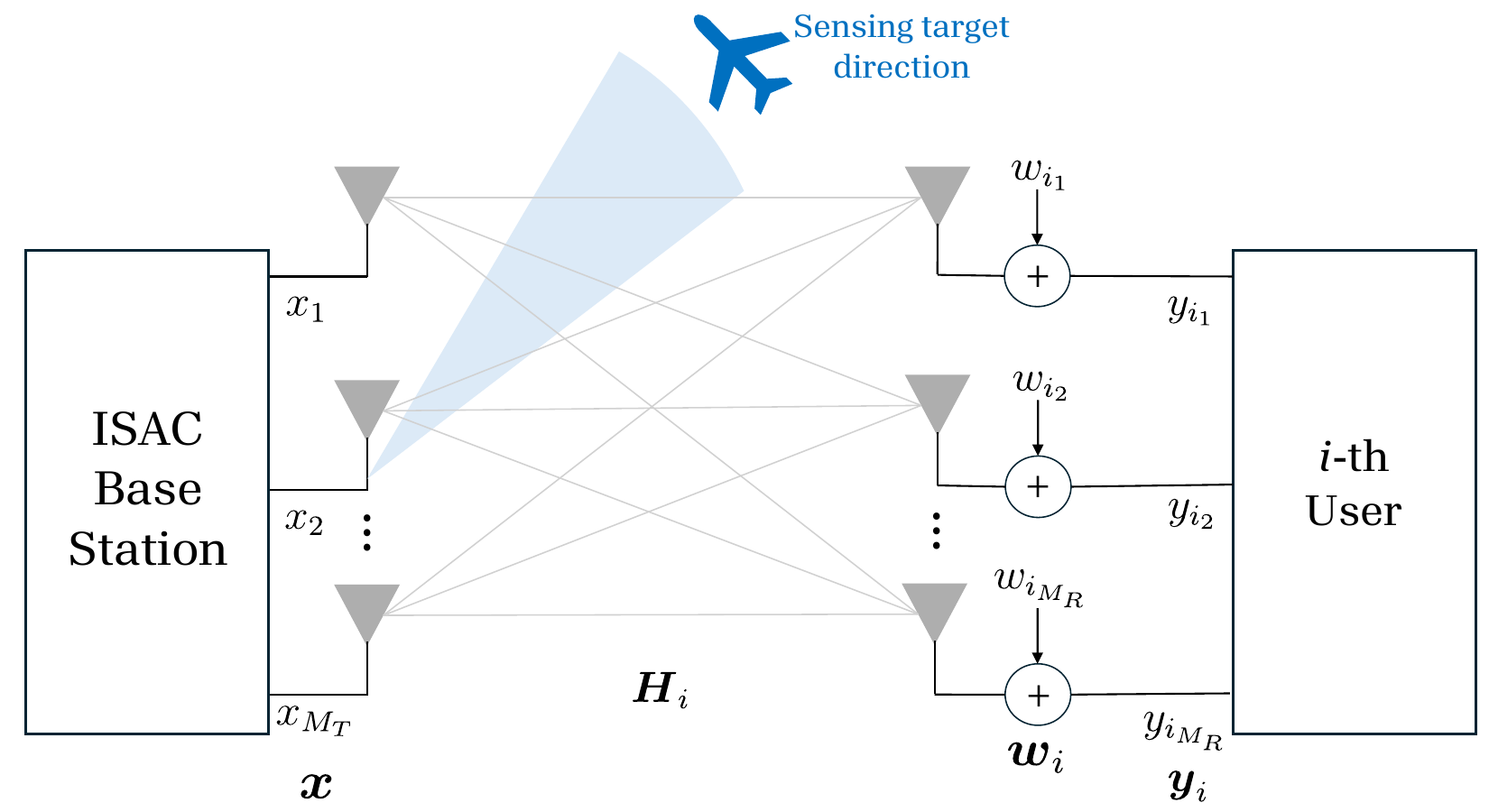}
    \caption{{Illustration of the considered system model.}}
    \label{system_model}
\end{figure}
We consider a {single-cell} \gls{isac} system employing a \gls{ula} with $M_T$ antennas and {half-wavelength inter-element spacing}. {The \gls{isac} \gls{bs} operates in the far-field and performs \gls{mimo} radar sensing while simultaneously serving $K_C < M_T$ users, where each user is equipped with $M_R$ antennas.

Let $\boldsymbol{x}[n] \in \mathbb{C}^{M_T \times 1}$ denote the baseband dual-functional \gls{tx} signal vector at the $n$-th discrete time instance within the block length of $N_t$, (i.e., $1 \le n \le N_t$), {used} to simultaneously perform sensing and communication. 

From a sensing perspective, the \gls{tx} beampattern describes how the radiated power is spatially distributed as a function of the azimuth direction~$\phi$, and can be expressed as \cite{li2008mimo}
\begin{equation}
    \label{Eqn:TransmitBeampattern}
    P(\phi)
    = \boldsymbol{a}_T^H(\phi) \boldsymbol{R}\boldsymbol{a}_T(\phi), 
\end{equation}
where $\boldsymbol{R} \coloneqq \mathbb{E}\left[\boldsymbol{x}[n]\boldsymbol{x}^H[n]\right]$ is the waveform covariance matrix, and $\boldsymbol{a}_T(\phi)$ denotes the \gls{tx} array steering vector, given by:
\begin{equation}
    \boldsymbol{a}_T(\phi) \coloneqq  [1~e^{j\pi \sin\phi}~...~e^{j(M_T-1) \pi \sin\phi }]^T.
\end{equation}

The \gls{tx} signal $\boldsymbol{x}[n]$ is also used for \gls{mumimo} communication
, and the corresponding complex baseband signal at the $i$-th communication user's \gls{rx} can be written as \cite{tse2005fundamentals} 
\begin{equation}
    \label{Eqn:SingleUser}
    \boldsymbol{y}_i[n] = \boldsymbol{H}_i \boldsymbol{x}[n] + \boldsymbol{w}_i[n], 
\end{equation}
where $\boldsymbol{H}_i \in \mathbb{C}^{M_R \times M_T}$ denotes the \gls{mimo} channel matrix between the \gls{bs} and $i$-th user, and $\boldsymbol{w}_i[n] \sim \mathcal{CN}(\boldsymbol{0}, \sigma_i^2\boldsymbol{I}_{M_R})$ are i.i.d. additive white Gaussian noise samples. {A simplified illustration of the considered system model is given in Fig. \ref{system_model}.}

{Although the beamforming framework proposed in this work is not constrained to any specific structure of channel matrix, for the analysis, we adopt the Rician channel model, given by \cite{tse2005fundamentals}} \looseness=-1
\begin{equation}
    \label{Eqn:RicianChannel}
    \boldsymbol{{H}}_i = \sqrt{\frac{\beta_i }{K+1}} \left( \sqrt{K}e^{j\psi_i}\boldsymbol{a}_R(\phi_i)\boldsymbol{a}_T^H(\phi_i) + \boldsymbol{{H}}_{sc} \right),
\end{equation}
where $K$ is the Rician factor, $\phi_i\in [-\pi/2,\pi/2]$ denotes the azimuth angle at which the \gls{bs} observes the user (i.e., communication \gls{los} direction), $\beta_i$ is the path-loss coefficient, $\psi_i \sim U[0,2\pi]$ is the uniformly distributed random phase, $\boldsymbol{a}_R(\phi_i) \in \mathbb{C}^{M_R\times 1}$ is the corresponding \gls{rx} array steering vector, and $\boldsymbol{H}_{sc}$ accounts for the multipath fading effects with the entries assumed to be i.i.d. with $\mathcal{CN}({0},1)$.  %

\section{Design Optimization Criteria}
\label{Sec:DesignOptCriteria}
\subsection{Beamforming Optimization for Sensing}
\label{Sec:SignalModelSensing}
The main objective of \gls{mimo} radar systems is to focus the available signaling energy towards the regions where radar targets are anticipated \cite{li2008mimo}. 
{To formulate this objective, we define the full grid $\Phi_g = \{ \bar{\phi}_i \}_{i=1}^{N_{\phi}}$, $N_{\phi} \gg M_T$, that ensures dense sampling for accurate beampattern approximation across all angles, with $\Phi_m \subseteq \Phi_g$ defining the mainlobe region where sensing targets are expected to be present.}
By assuming that every $\bar{\phi}_i \in \Phi_m$ is given the same priority, the ideal normalized \gls{tx} beampattern is defined as 
\begin{equation}
    \label{Eqn:DesiredBeampatternDefn}
    P_d(\bar{\phi}_i) \coloneqq 
    \begin{cases}
        1,~ &\mathrm{if}~ \bar{\phi}_i\in \Phi_m, \\
        \varepsilon,~& \mathrm{otherwise}, 
    \end{cases}
\end{equation}
where $\varepsilon \ge 0$ is a sufficiently small constant, corresponding to the desired \gls{psl}. 

The objective of \gls{mimo} radar is to construct $\boldsymbol{R}$ such that $P(\bar{\phi}_i)$ {is as close as possible to} $\eta P_d(\bar{\phi}_i)$, for every $\bar{\phi}_i \in \Phi_g$, where $\eta > 0$ is introduced for power normalization since $P({\phi})$, as seen in \eqref{Eqn:TransmitBeampattern}, scales with \gls{tx} power. 
At the same time, for improved radar \gls{rx} processing, the cross-correlations between signals transmitted in different directions should be low. This implies that $|\boldsymbol{a}_T^H(\bar{\phi}_i)\boldsymbol{R}\boldsymbol{a}_T(\bar{\phi}_j)|$ should be sufficiently small for $\bar{\phi}_i \ne \bar{\phi}_j$ \cite{stoica2007probing}. 
In our recent work \cite{globecom_bk}, we unified these two objectives based on the following \textit{matrix nearness} formulation
\begin{equation}
    \label{Eqn:CostSensing}
    J_s(\boldsymbol{R}) \coloneqq \Vert \boldsymbol{D}^H \boldsymbol{R} \boldsymbol{D} - \eta \boldsymbol{T} \Vert_F^2,   
\end{equation}
where the columns of $\boldsymbol{D}$ contain steering vectors corresponding to all grid points (array manifold), i.e., 

\begin{align}
    \label{Eq:D}
    \boldsymbol{D} \coloneqq {\frac{1}{\sqrt{N_{\phi}}}}\left[\boldsymbol{a}_T(\bar{\phi}_1)~ \boldsymbol{a}_T(\bar{\phi}_2)~...~\boldsymbol{a}_T(\bar{\phi}_{N_{\phi}})\right],
\end{align} 
and
\begin{equation}
\label{Eqn:Tdefn}
T_{ij} \coloneqq 
\begin{cases}
    P_d(\Bar{\phi}_i), &\mathrm{if~} i=j, \\
    0,~&\mathrm{otherwise},
\end{cases}
\end{equation}
{whose diagonals determine the desired beampattern, with zero off-diagonals introduced to suppress cross-correlations.}
When the grid points $\Bar{\phi}_i$ are obtained by uniformly sampling the electrical angle domain, $\pi \sin \phi \in[-\pi, \pi)$, for the considered \gls{ula} configuration, {$\boldsymbol{D}\boldsymbol{D}^H = \boldsymbol{I}_{M_T}$ holds and, as shown in our prior work \cite{globecom_bk}}, we can rewrite \eqref{Eqn:CostSensing} as 
\begin{align}
\label{Eqn:CostSensing2}
J_s(\boldsymbol{R})
    =\Vert \boldsymbol{{{R}}} - \boldsymbol{\Tilde{T}}\Vert_F^2 + {C},
\end{align}
where $\boldsymbol{\Tilde{T}} \coloneqq \eta\boldsymbol{D}\boldsymbol{{T}}\boldsymbol{D}^H$, 
and the term ${C}$ does not depend on $\boldsymbol{R}$ {and is therefore irrelevant to the optimization over $\boldsymbol{R}$.} 

\subsection{Beamforming Optimization for Communication}
\label{Sec:SignalModelCommunication}
We adopt the \gls{snr} {experienced by each user} as the communication \gls{qos} metric \cite{gershman2010convex}.
By {considering} the standard assumption of independent $\boldsymbol{H}_i$ and $\boldsymbol{x}$ \cite{telatar1999capacity}, the \gls{snr} experienced by the $i$-th user can be written as\footnote{{The multi-user interference is not explicitly considered here; its suppression is assumed to be addressed in the successive waveform design step \cite{kilic_jsac}.}}
\begin{equation}
    \label{Eqn:SNRExpression}
    \mathrm{SNR}_i = \frac{{E[\Vert \boldsymbol{H}_i\boldsymbol{x} \Vert^2]}}{E[\Vert \boldsymbol{w}_i \Vert^2]} = \frac{\mathrm{Tr}(\boldsymbol{\Omega}_i \boldsymbol{R})}{\sigma_i^2 M_R},  
\end{equation}
where $\boldsymbol{\Omega}_i \coloneqq \mathbb{E}[\boldsymbol{H}_i^H \boldsymbol{H}_i]$ is the covariance matrix of the channel, {which} is assumed to be known by the \gls{bs}. Since $\boldsymbol{\Omega}_i$ changes much slower than the fast fading channel matrix $\boldsymbol{H}_i$, acquiring the knowledge of $\boldsymbol{\Omega}_i$ entails a significantly lower overhead than the \gls{csi} feedback of individual channel realizations $\boldsymbol{H}_i$. 
The resulting approach is often called long-term beamforming \cite{lozano2007long}. 
Imposing the \gls{snr} requirements means that the following inequalities need to be satisfied
\begin{equation}
    \label{Eqn:CommConstraint}
    \mathrm{Tr}(\boldsymbol{\Omega}_i\boldsymbol{R}) \ge \Tilde{\Gamma}_i\sigma_i^2 M_R, ~ 1 \le i \le K_C,
\end{equation}
where $\Tilde{\Gamma}_i \ge 0$ is the minimum tolerable \gls{snr} for the $i$-th user.

\section{Waveform Covariance Design Methodology}
\label{Sec:ProposedMethodology}

\subsection{Proposed Problem Formulation}
Following the design criteria described in Sections \ref{Sec:SignalModelSensing} and \ref{Sec:SignalModelCommunication}, we can formulate an optimization problem admitting \eqref{Eqn:CostSensing2} as the sensing cost function and including communication constraints \eqref{Eqn:CommConstraint}. 
For practical feasibility, $\boldsymbol{R}$ must also satisfy additional conditions. Let $P_T$ denote the total power budget of the system. 
Assuming that the energy budget is to be fully utilized ($\mathrm{Tr}(\boldsymbol{R})=P_T$), under the equal per-antenna power constraint, we must have
$R_{jj}=P_T/M_T$, $1 \le j \le M_T$.  Moreover, as a covariance matrix, $\boldsymbol{R}$ must be positive semidefinite by definition. 

We can now formulate a convex optimization framework for waveform covariance matrix design in \gls{isac} systems:
\begin{align}
    \label{Eqn:ProposedOptCovMtx2}
    \min_{\boldsymbol{R}}&~\frac{1}{2}\Vert \boldsymbol{R} - \boldsymbol{\Tilde{T}} \Vert_F^2 \nonumber \\
    \mathrm{s.t.~} & \mathrm{Tr}(\boldsymbol{\Omega}_i\boldsymbol{R}) \ge \Gamma_i, ~ 1\le i \le K_C, \nonumber \\
    & \mathrm{Tr}(\boldsymbol{E}_j \boldsymbol{R}) = P_T/M_T, ~ 1\le j \le M_T, \nonumber \\ 
    &\boldsymbol{R} \succeq \boldsymbol{0},   
\end{align}
where $\Gamma_i \coloneqq \Tilde{\Gamma}_i\sigma_i^2 M_R$ and $\boldsymbol{E}_j$ is a matrix with $1$ at its $j$-th diagonal and zeros otherwise. We note that by removing the communication constraints from \eqref{Eqn:ProposedOptCovMtx2}, we obtain our previously proposed sensing covariance matrix design approach \cite{globecom_bk}. \looseness=-1

\subsection{Solution via Direct Eigendecomposition}
Assuming that the Slater's condition holds, i.e., there exists an $\boldsymbol{R} \succ \boldsymbol{0}$ satisfying the affine constraints in \eqref{Eqn:ProposedOptCovMtx2}, strong duality holds and the problem can be optimally solved via its dual. Here we extend our dual problem formulation from \cite{globecom_bk} by additionally considering the communication constraints. As these constraints are affine, the extension is straightforward. The reader is referred to \cite{globecom_bk, kilic_jsac} for the full derivation. The solution is based on the approach proposed in \cite{boyd2005least}, accelerated with adaptively restarted momentum terms \cite{beck2009fast, giselsson2014monotonicity, o2015adaptive} for improved convergence. Its outline is provided as follows. \looseness=-1

Let $\mu_j \in \mathbb{R}$, and $\nu_i \ge 0$ be the Lagrange multipliers associated with the affine equality and inequality constraints, respectively. The corresponding terms with momentum updates are denoted by $\Bar{\mu}_j$ and $\Bar{\nu}_i$. 
If $\boldsymbol{\mu}^\star$ and $\boldsymbol{\nu}^\star$ are the optimal Lagrange multipliers, which can be found according to Algorithm \ref{Algo:ProjectedGradient}, then the optimal covariance matrix is $\boldsymbol{R}^{\star} \coloneqq \boldsymbol{R}(\boldsymbol{\mu}^\star, \boldsymbol{\nu}^\star)$ where $\boldsymbol{R}$ at the $k$-th iteration is computed as  

\begin{equation}
    \label{Eqn:OptCov}
    \boldsymbol{R}(\Bar{\boldsymbol{\mu}}^k, \Bar{\boldsymbol{\nu}}^k) = \left[\boldsymbol{\Tilde{T}} - \boldsymbol{B}(\Bar{\boldsymbol{\mu}}^k) + \boldsymbol{C}(\Bar{\boldsymbol{\nu}}^k) \right]_+, 
\end{equation}
with
\begin{equation}
    \label{Eqn:EmuDefn}
    \boldsymbol{B}(\Bar{\boldsymbol{\mu}}^k) \coloneqq \sum_{j=1}^{M_T} \Bar{\mu}_j^k \boldsymbol{E}_j,
\end{equation}
\begin{equation}
\boldsymbol{C}(\Bar{\boldsymbol{\nu}}^k) \coloneqq \sum_{i=1}^{K_C} \Bar{\nu}_i^k \boldsymbol{\Omega}_i.
\end{equation}
In \eqref{Eqn:OptCov}, the operator $[\boldsymbol{M}]_+$ projects a Hermitian matrix $\boldsymbol{M}$ onto the positive semidefinite cone, i.e, 
\begin{equation}
    [\boldsymbol{M}]_+ \coloneqq \sum_{i} (\lambda_i)_+ \boldsymbol{q}_i \boldsymbol{q}_i^H, 
\end{equation}
where $\lambda_i$ and $\boldsymbol{q}_i$ denote the eigenvalues and eigenvectors of $\boldsymbol{M}$, respectively, and $(\lambda_i)_+ \coloneqq \max(\lambda_i, 0)$.

The evaluation of \eqref{Eqn:OptCov} is required for the gradient computations in Steps 6 and 7 (and also in Steps 10 and 11 if the restart condition holds) of Algorithm \ref{Algo:ProjectedGradient}. In these steps, $\gamma=1/L$ denotes the step-size, where $L$ is the Lipschitz constant for the mapping from ($\boldsymbol{\nu, \mu}$) to these gradients, which can be computed as \cite{boyd2005least} 
\begin{equation}
    L = \sum_{j=1}^{M_T}\lambda^2_{\mathrm{max}}(\boldsymbol{E}_j) + \sum_{i=1}^{K_C}\lambda^2_{\mathrm{max}}(\boldsymbol{\Omega}_i) = M_T + \sum_{i=1}^{K_C}\lambda^2_{\mathrm{max}}(\boldsymbol{\Omega}_i), 
\end{equation} 
where $\lambda_{\mathrm{max}}(.)$ denotes the maximum eigenvalue.

\begin{algorithm}[t]
	\caption{Procedure to solve \eqref{Eqn:ProposedOptCovMtx2}, based on accelerated projected gradient method with adaptive momentum restart}
	\begin{algorithmic}[1] 
	\label{Algo:ProjectedGradient}
		\renewcommand{\algorithmicrequire}{\textbf{Input:}}
		\renewcommand{\algorithmicensure}{\textbf{Output:}}
        \REQUIRE $\boldsymbol{\Tilde{T}}$, $\{ \boldsymbol{\Omega}_i \}_{i=1}^{K_C}$, $\{ \boldsymbol{\Gamma}_i \}_{i=1}^{K_C}$, $P_T$, $M_T$
		\ENSURE $\boldsymbol{\mu}^\star$, $\boldsymbol{\nu}^\star$
        \STATE {Initialize} $\boldsymbol{\nu}^0=\boldsymbol{{\nu}}^{-1}=\boldsymbol{0}$, $\boldsymbol{\mu}^0=\boldsymbol{\mu}^{-1}=\boldsymbol{0}$, $t^0= t^{-1} = 1$ 
        \FOR{$k\ge 0$} 
        \STATE $t^{k}=\big(1+\sqrt{1+4(t^{k-1})^2}\big)/2$
        \STATE $\boldsymbol{\Bar{\nu}}^{k} = \boldsymbol{{\nu}}^{k}+\big((t^{k-1}-1)/t^{k}\big)(\boldsymbol{{\nu}}^{k}-\boldsymbol{{\nu}}^{k-1})$
        \STATE $\boldsymbol{\Bar{\mu}}^{k} = \boldsymbol{{\mu}}^{k}+\big((t^{k-1}-1)/t^{k}\big)({\boldsymbol{\mu}}^{k}-{\boldsymbol{\mu}}^{k-1})$
        \STATE ${{\nu}}_i^{k+1} = \left( {\Bar{\nu}}_i^{k} + \gamma (-\mathrm{Tr}(\boldsymbol{\Omega}_i\boldsymbol{R}(\boldsymbol{\Bar{\mu}}^k, \boldsymbol{\Bar{\nu}}^k)) + \Gamma_i) \right)_+$ \vspace{1pt}
        \STATE ${{\mu}}_j^{k+1} =  {\Bar{\mu}}^{k}_j + \gamma (\mathrm{Tr}(\boldsymbol{E}_j\boldsymbol{R}(\boldsymbol{\Bar{\mu}}^k, \boldsymbol{\Bar{\nu}}^k))-P_T/M_T) $ \vspace{1pt}
        \IF{$(\boldsymbol{\Bar{\nu}}^k-\boldsymbol{\nu}^{k+1})^T(\boldsymbol{\nu}^{k+1}-\boldsymbol{\nu}^k)
        +({\boldsymbol{\Bar{\mu}}}^k-\boldsymbol{{\mu}}^{k+1})^T(\boldsymbol{\mu}^{k+1}-{\boldsymbol{\mu}}^k) > 0$}
        \STATE $\boldsymbol{\Bar{\nu}}^{k}=\boldsymbol{{\nu}}^{k}$, $\boldsymbol{\Bar{\mu}}^{k}=\boldsymbol{{\mu}}^{k}$
        \STATE ${{\nu}}_i^{k+1} = \left( {\Bar{\nu}}_i^{k} + \gamma (-\mathrm{Tr}(\boldsymbol{\Omega}_i\boldsymbol{R}(\boldsymbol{\Bar{\mu}}^k, \boldsymbol{\Bar{\nu}}^k)) + \Gamma_i) \right)_+$ \vspace{1pt}
        \STATE ${{\mu}}_j^{k+1} =  {\Bar{\mu}}^{k}_j + \gamma (\mathrm{Tr}(\boldsymbol{E}_j\boldsymbol{R}(\boldsymbol{\Bar{\mu}}^k, \boldsymbol{\Bar{\nu}}^k))-P_T/M_T) $ \vspace{1pt}
        \ENDIF
        \IF {$\Vert \boldsymbol{\bar{\nu}}^{k} - \boldsymbol{\nu}^{k+1} \Vert/K_C +\Vert \boldsymbol{\bar{\mu}}^{k} - \boldsymbol{\mu}^{k+1} \Vert/M_T
    \le \epsilon$}
    \STATE {report $\boldsymbol{\mu}^\star = \boldsymbol{\mu}^{k+1}$ and $\boldsymbol{\nu}^\star=\boldsymbol{\nu}^{k+1}$} 
    \ENDIF
        \ENDFOR
	\end{algorithmic} 
    
\end{algorithm}

\subsection{Derivation of a Condition to Avoid Eigendecomposition}
\label{sec:accel}
The computationally dominant operation in Algorithm \ref{Algo:ProjectedGradient} is the evaluation of the \gls{evd} as part of the projection in \eqref{Eqn:OptCov}, with complexity of $O(M_T^3)$. While Algorithm~\ref{Algo:ProjectedGradient} remains significantly more efficient than alternative interior-point methods, which have a per-iteration complexity of $O(M_T^6)$ when applied to solve \eqref{Eqn:ProposedOptCovMtx2}~\cite{boyd2005least},
its computational load can be further reduced by avoiding the projection step if the argument of $[.]_+$ in \eqref{Eqn:OptCov} is already positive semidefinite, i.e.,
\begin{equation}
    \label{Eqn:PosSemDefCheck}
    \lambda_{\mathrm{min}}\left(\boldsymbol{\Tilde{T}} - \boldsymbol{B}(\Bar{\boldsymbol{\mu}}^k) + \boldsymbol{C}(\Bar{\boldsymbol{\nu}}^k)\right) \ge 0,
\end{equation}
where $\lambda_{\mathrm{min}}(.)$ denotes the minimum eigenvalue. 
Here, we derive a simple condition to verify \eqref{Eqn:PosSemDefCheck} to avoid the unnecessary \gls{evd} computations.

For Hermitian matrices $\boldsymbol{A}_1$ and $\boldsymbol{A}_2$, according to the Courant-Fischer theorem \cite[p.~783,~Thm.~20.A.1]{marshall1979inequalities}, we have 
\begin{equation}
    \label{Eqn:CourantFischer}
    \lambda_{\mathrm{min}}(\boldsymbol{A}_1+\boldsymbol{A}_2)\ge \lambda_{\mathrm{min}}(\boldsymbol{A}_1) + \lambda_{\mathrm{min}}(\boldsymbol{A}_2).  
\end{equation}
Since $\boldsymbol{\Tilde{T}}$, $\boldsymbol{B}(\boldsymbol{\Bar{\mu}}^k)$ and $\boldsymbol{C}(\boldsymbol{\Bar{\nu}}^k)$ are all Hermitian matrices, we can apply \eqref{Eqn:CourantFischer} to \eqref{Eqn:PosSemDefCheck}, i.e., 
\begin{align}
    \label{Eqn:CondConstrained}
    {\lambda}_{\mathrm{min}}\left(\boldsymbol{\Tilde{T}} - \boldsymbol{B}(\boldsymbol{\Bar{\mu}}^k) + \boldsymbol{C}(\boldsymbol{\Bar{\nu}}^k)\right) \nonumber \\ \ge 
    {\lambda}_{\mathrm{min}}(\boldsymbol{\Tilde{T}})+{\lambda}_{\mathrm{min}}( -\boldsymbol{B}(\boldsymbol{\Bar{\mu}}^k)) +{\lambda}_{\mathrm{min}}( \boldsymbol{C}(\boldsymbol{\Bar{\nu}}^k)). 
\end{align}
Since
$\boldsymbol{B}(\boldsymbol{\Bar{\mu}}^k)$ is a diagonal matrix by definition \eqref{Eqn:EmuDefn}, we write 
\begin{equation}
    \label{Eqn:CourantFischerMu}
    {\lambda}_{\mathrm{min}}( -\boldsymbol{B}(\boldsymbol{\Bar{\mu}}^k)) = -{\lambda}_{\mathrm{max}}( \boldsymbol{B}(\boldsymbol{\Bar{\mu}}^k)) = -\max_i \Bar{\mu}_i^k.
\end{equation}
On the other hand, since $\Bar{\nu}_i^k \ge 0$ and ${\lambda}_{\mathrm{min}} (\boldsymbol{\Omega}_i) \ge 0$, from \eqref{Eqn:CourantFischer} we have, 
\begin{align}
    \label{Eqn:CourantFischerNu}
    {\lambda}_{\mathrm{min}}( \boldsymbol{C}(\boldsymbol{\Bar{\nu}}^k)) ={\lambda}_{\mathrm{min}}\left(\sum_{i=1}^{K_C} \Bar{\nu}_i^k \boldsymbol{\Omega}_i\right)  \ge
    \sum_{i=1}^{K_C} \Bar{\nu}_i^k {\lambda}_{\mathrm{min}} (\boldsymbol{\Omega}_i). 
\end{align}
By combining \eqref{Eqn:CondConstrained}, \eqref{Eqn:CourantFischerMu} and \eqref{Eqn:CourantFischerNu}, we obtain
\begin{align}
    \label{Eqn:CondConstrained2}
    &{\lambda}_{\mathrm{min}}\left(\boldsymbol{\Tilde{T}} - \boldsymbol{B}(\boldsymbol{\Bar{\mu}}^k) + \boldsymbol{C}(\boldsymbol{\Bar{\nu}}^k)\right) \nonumber \\ 
    &\ge {\lambda}_{\mathrm{min}}(\boldsymbol{\Tilde{T}})- \max_j \Bar{\mu}_j^k + \sum_{i=1}^{K_C} \Bar{\nu}_i^k  {\lambda}_{\mathrm{min}} (\boldsymbol{\Omega}_i).
\end{align}

By combining \eqref{Eqn:PosSemDefCheck} and \eqref{Eqn:CondConstrained2}, we finally obtain a sufficient condition to avoid the \gls{evd} computation in \eqref{Eqn:OptCov}, i.e., 
\begin{equation}
    \label{Eqn:CheckPosSemDef}
    {\lambda}_{\mathrm{min}}(\boldsymbol{\Tilde{T}})- \max_j \Bar{\mu}_j^k + \sum_{i=1}^{K_C} \Bar{\nu}_i^k  {\lambda}_{\mathrm{min}} (\boldsymbol{\Omega}_i) \ge 0,
\end{equation}
which guarantees the positive semidefiniteness of the argument of $[.]_+$ in  \eqref{Eqn:OptCov}. 
Moreover, ${{\lambda}_{\mathrm{min}}(\boldsymbol{\Tilde{T}})}$ and ${\lambda}_{\mathrm{min}} (\boldsymbol{\Omega}_i)$ need to be computed only once, therefore, their computation does not increase the computational complexity of Algorithm \ref{Algo:ProjectedGradient}. 

\section{Simulation Results}
\label{Sec:NumResults}
\subsection{Considered Simulation Scenario and Performance Metrics}
To evaluate the performance of the proposed method, we consider a long-term beamforming scenario under the Rician channel model, i.e., $\boldsymbol{\Omega}_i=\mathbb{E}[\boldsymbol{H}_i^H \boldsymbol{H}_i]$ where $\boldsymbol{H}_i$ is modeled according to \eqref{Eqn:RicianChannel} with Rician factor $K=5$.  
{We simulate $K_C = 5$ communication users with \gls{los} directions $\{-60^{\circ}, -40^{\circ}, 0^{\circ}, 30^{\circ}, 55^{\circ}\}$. The numbers of antennas at the \gls{bs} and at each user are set to $M_T = 128$ and $M_R = 2$, respectively.}
The path loss and noise standard deviation parameters of all users are fixed to unity, i.e., $\beta_i=  \sigma_i= 1$ for all $1 \le i \le K_C$. The total \gls{bs} power budget is $P_T = 43$ dBm. The grid size is $N_{\phi}=2048$. 
{Although the method supports different \gls{snr} thresholds per user, we set $\Tilde{\Gamma}_i = \Tilde{\Gamma}$ for all $1 \le i \le K_C$ for simplicity.}
To generate the desired beampattern, we set $\varepsilon=0.01$ in \eqref{Eqn:DesiredBeampatternDefn}. We fix $\epsilon=10^{-10}$ in Step 12 of Algorithm \ref{Algo:ProjectedGradient} for the stopping condition. {We set $\eta=P_T/\mathrm{Tr}(\boldsymbol{D}\boldsymbol{T}\boldsymbol{D}^H)$ in (9), resulting in $\mathrm{Tr}(\boldsymbol{\Tilde{T}}) =  \mathrm{Tr}(\boldsymbol{R})$.}

We consider a scenario with a single sensing mainlobe located at $0^{\circ}$, with a desired beamwidth of $20^{\circ}$. To evaluate the beampattern quality for sensing, we calculate the \gls{psl} and the fraction of the total \gls{tx} power transmitted towards mainlobe directions, i.e., $P_m/P_T$ \cite{globecom_bk}. 
In Section \ref{Sec:SensCommTradeoff}, we investigate the sensing performance of the beampatterns obtained based on the framework \eqref{Eqn:ProposedOptCovMtx2} under different communication \gls{snr} constraints. In Section \ref{Sec:AvoidingEig}, we analyze the effectiveness of the conditioned acceleration method from Section \ref{sec:accel}. 

\subsection{Sensing and Communication Trade-off}
We examine the sensing performance under different \gls{snr} requirements for communication users by changing $\Tilde{\Gamma}$ in \eqref{Eqn:CommConstraint}. 
Fig. \ref{comm_sens_tradeoff} shows that when higher \glspl{snr} are required, the obtained $P_m/P_T$ values decrease and the \gls{psl} increases, i.e., the sensing performance deteriorates. This is expected since more power for communication must be allocated to meet higher communication \gls{snr} requirements.
\begin{figure}[t]
    \centering
    \includegraphics[width=\linewidth]{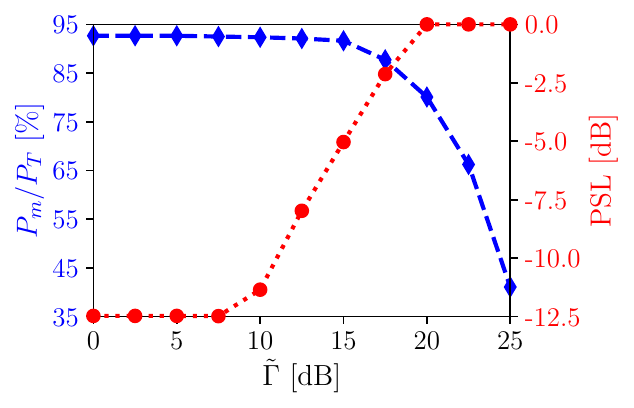} 
    \caption{Sensing performance under different communication \gls{snr} constraints.}
    \label{comm_sens_tradeoff}
\end{figure}
\label{Sec:SensCommTradeoff}

This can also be observed from Fig. \ref{bp_snrs}, showing the beampatterns generated for $\Tilde{\Gamma}=10$ dB and $\Tilde{\Gamma}=25$ dB. 
The figure clearly shows that stronger beams are formed in the \gls{los} directions of users for the minimum required \gls{snr} of 25 dB. For the \gls{los} direction at $\phi = 0^{\circ}$, which is in the sensing region of the mainlobe, for $\Tilde{\Gamma}=25$ dB, we observe a peak, which is not present when $\Tilde{\Gamma}=10$ dB, since in the latter case the sensing power along $\phi = 0^{\circ}$ already satisfies the \gls{snr} constraint for the corresponding user.  
\begin{figure}[t]
    \centering
    \includegraphics[width=0.9\linewidth]{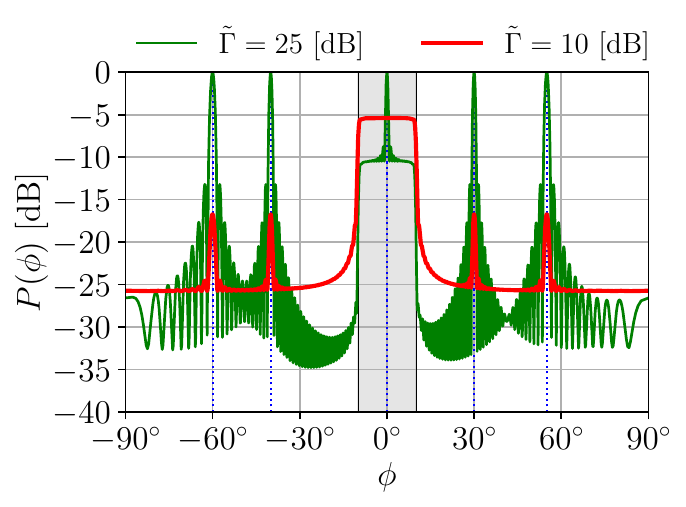} 
    \caption{Normalized beampatterns under communication constraints $\Tilde{\Gamma}=10$ dB and $\Tilde{\Gamma}=25$ dB. Blue lines denote the communication \gls{los} directions and the gray region denotes the desired sensing beampattern. }
    \label{bp_snrs}
\end{figure}

\subsection{The Effectiveness of Algorithm Acceleration} 
\label{Sec:AvoidingEig}
To analyze the effect of implementing the condition \eqref{Eqn:CheckPosSemDef}, Table \ref{tab:snr_comparison} summarizes the computation times obtained with the Conditioned Algorithm \ref{Algo:ProjectedGradient} (C-Alg. 1), i.e., implementing the acceleration method described in Section \ref{sec:accel}, together with those obtained with the (non-accelerated) Algorithm \ref{Algo:ProjectedGradient}. We report results for $\Tilde{\Gamma}$ between 10\,dB and 25\,dB, with 2.5\,dB increment.
The reported execution times are evaluated on a personal computer, with an Intel i7-1365U CPU and 16 GB RAM. 
As we observe from the table, C-Alg. 1 avoids all \gls{evd} computations\footnote{The number of iterations required for convergence together with the number of momentum restarts give the total number of \gls{evd} evaluations.} for $\Tilde{\Gamma} \le 15$ dB. For these scenarios, C-Alg. 1 reduces the computation time by around six times, where it takes only around $2$ s to get the final covariance matrix output.  

When $\Tilde{\Gamma}$ increases, the effectiveness of the acceleration method in C-Alg. 1 reduces. For $\Tilde{\Gamma} = 17.5$ dB, C-Alg. 1 avoids $30\%$ of the required \gls{evd} computations, while this ratio is $19\%$, $14\%$, and $11\%$ when  $\Tilde{\Gamma} = 20$ dB, $\Tilde{\Gamma} = 22.5$ dB, and $\Tilde{\Gamma} = 25$ dB, respectively. However, C-Alg. 1 always reduces the number of required \gls{evd} evaluations and hence the computation time.  
Interestingly, C-Alg. 1 successfully avoids all unnecessary \glspl{evd} in all scenarios considered here, i.e., the condition \eqref{Eqn:CheckPosSemDef} successfully detects all the cases where \eqref{Eqn:PosSemDefCheck} holds. \looseness-1

Finally, we note that although \eqref{Eqn:ProposedOptCovMtx2} is a convex optimization problem and can, in principle, be solved using general-purpose solvers, practical implementation poses significant challenges. Specifically, when the problem is formulated in CVXPY~\cite{diamond2016cvxpy} and solved using the modern interior-point solver Clarabel~\cite{Clarabel_2024}, the solver fails on the personal computer used for this analysis, due to an excessive memory consumption caused by the high problem dimensionality.

\begin{table}[t]
\caption{Comparison of computation times and the number of required \gls{evd} evaluations across different \gls{snr} values.}
\centering
\begin{tabular}{|l|ccccccc|}
\hline
\multicolumn{1}{|c|}{} & \multicolumn{7}{c|}{SNR [dB]} \\
\cline{2-8}
 & 10.0 & 12.5 & 15.0 & 17.5 & 20.0 & 22.5 & 25.0 \\
\hline
\multicolumn{8}{|c|}{{Elapsed Time [s]}} \\
\hline
Alg. 1 & 12.59 & 13.77 & 14.86 & 49.74 & 61.12 & 65.55 & 73.77 \\
C-Alg. 1 & 2.16 & 2.37 & 2.44 & 39.02 & 52.06 & 57.93 & 63.96 \\
\hline
\multicolumn{8}{|c|}{{Number of Required \gls{evd} Evaluations}} \\
\hline
Alg. 1 & 1620 & 1756 & 1812 & 4085 & 4996 & 5366 & 5738 \\
C-Alg. 1 & 0 & 0 & 0 & 2841 & 4044 & 4593 & 5057 \\
\hline
\end{tabular}
\label{tab:snr_comparison}
\end{table}

\section{Conclusions}
\label{Sec:Conclusions}
In this letter, we propose a novel \gls{isac} beamforming method for joint \gls{mimo} radar sensing and \gls{mumimo} communication. The contribution of this work is twofold.
First, we extend our previous beamforming problem formulation \cite{globecom_bk} for \gls{isac} systems, by incorporating additional communication \gls{snr} constraints.
Second, we propose an acceleration method to reduce the required computation time, by avoiding the time-consuming \gls{evd} evaluations playing the central role in the solution algorithm.
Unlike most of the existing approaches, our method only requires \gls{csi} statistics at \gls{bs}, thereby enhancing its practical feasibility and reducing the overhead of \gls{csi} feedback required from users to \gls{bs}.

Based on numerical simulations, we demonstrate that the proposed method can synthesize beampatterns meeting both communication and sensing requirements. Moreover, the proposed acceleration method is observed to reduce the overall computation time by up to six times, by effectively avoiding unnecessary \gls{evd} evaluations in the considered scenarios.

\bibliographystyle{IEEEtran}
\bibliography{ieeeTran_bibliography_isac}

\end{document}

%% file: utils/acronyms.tex
\newacronym{tx}{Tx}{transmitter}
\newacronym{rx}{Rx}{receiver}
%
\newacronym{mimo}{MIMO}{multiple-input multiple-output}
\newacronym{mumimo}{MU-MIMO}{multi-user MIMO}
\newacronym{ula}{ULA}{uniform linear array}
%
\newacronym{fista}{FISTA}{Fast Iterative Shrinkage-Thresholding Algorithm}
\newacronym{pg}{PG}{projected gradient}
\newacronym{apg}{APG}{accelerated PG}
\newacronym{am}{AM}{alternating minimization}
%
\newacronym{rcs}{RCS}{radar cross-section}
\newacronym{isac}{ISAC}{integrated sensing and communication}
\newacronym{psl}{PSL}{peak sidelobe level}

\newacronym{bpm}{BPM}{beampattern matching}
\newacronym{svd}{SVD}{singular value decomposition}

\newacronym{snr}{SNR}{signal-to-noise ratio}
\newacronym{sinr}{SINR}{signal-to-interference-plus-noise ratio}

\newacronym{mui}{MUI}{multi-user interference}
\newacronym{pa}{PA}{power amplifier}
\newacronym{rf}{RF}{radio-frequency}
\newacronym{par}{PAPR}{peak-to-average-power-ratio}
\newacronym{cm}{CM}{constant-modulus}
\newacronym{qpsk}{QPSK}{quadrature phase shift keying}

\newacronym{mse}{MSE}{mean square error}
\newacronym{wmse}{MSE}{weighted mean square error}

\newacronym{ls}{LS}{least squares}
\newacronym{wls}{WLS}{weighted least squares}

\newacronym{mmse}{MMSE}{minimum mean square error}
\newacronym{zf}{ZF}{zero-forcing}
\newacronym{ser}{SER}{symbol error rate}
\newacronym{mf}{MF}{matched filter}

\newacronym{mrt}{MRT}{maximum ratio transmission}

\newacronym{bs}{BS}{base station}
\newacronym{los}{LoS}{line-of-sight}
\newacronym{dac}{DAC}{digital-to-analog converter}
\newacronym{qsdp}{QSDP}{quadratic semidefinite programming problem}
\newacronym{wf}{WF}{Wiener filter}
\newacronym{csi}{CSI}{channel state information}
\newacronym{qam}{QAM}{quadrature amplitude modulation}
\newacronym{doa}{DOA}{direction-of-arrival}
\newacronym{crb}{CRB}{Cramer-Rao Bound}
\newacronym{lfm}{LFM}{linear frequency modulation}
\newacronym{sdr}{SDR}{semi-definite relaxation}

\newacronym{ismr}{ISMR}{integrated sidelobe to mainlobe ratio}

\newacronym{tdd}{TDD}{time-division-duplex}
\newacronym{dft}{DFT}{discrete Fourier transform}
\newacronym{apes}{APES}{amplitude and phase estimation}
\newacronym{sir}{SIR}{signal-to-interference ratio}
\newacronym{is}{IS}{interference suppression}
\newacronym{qos}{QoS}{quality-of-service}
\newacronym{evd}{EVD}{eigenvalue decomposition}
\newacronym{fdd}{FDD}{frequency division duplexing}